\documentclass{appolb}
\usepackage{graphicx}

\usepackage[T1]{fontenc}
\usepackage[utf8]{inputenc}  
\usepackage{amsmath}
\usepackage{cite}     
\usepackage{hyperref} 

\usepackage{multirow}
\usepackage{multicol}

\begin{document}
\title{Analysis procedure of the positronium lifetime spectra for the J-PET detector}%

\author{
K.~Dulski$^c$\footnote{Corresponding author: kamil.dulski@gmail.com}, B.~Zgardzi\'nska$^e$,
P.~Bia\l{}as$^c$, C.~Curceanu$^d$ E.~Czerwi\'nski$^c$, A.~Gajos$^c$, B.~G\l{}owacz$^c$, M.~Gorgol$^e$, B.~C.~Hiesmayr$^f$, B.~Jasi\'nska$^e$, 
D.~Kisielewska-Kami\'nska$^c$, G.~Korcyl$^c$, P.~Kowalski$^{a}$, T.~Kozik$^c$, N.~Krawczyk$^c$, W.~Krzemie\'n$^{b}$, E.~Kubicz$^c$, M.~Mohammed$^{c,g}$, 
M.~Pawlik-Nied\'zwiecka$^c$, S.~Nied\'zwiecki$^c$, M.~Pa\l{}ka$^c$, L.~Raczy\'nski$^{a}$, J.~Raj$^c$, Z.~Rudy$^c$, N.~G.~Sharma$^c$, S.~Sharma$^c$, Shivani$^c$, R.~Y.~Shopa$^{a}$, M.~Silarski$^c$, M.~Skurzok$^c$, 
A.~Wieczorek$^c$, W.~Wi\'slicki$^{a}$, M.~Zieli\'nski$^c$, P.~Moskal$^c$
\address{
$^{a}$ Department of Complex Systems, National Centre for Nuclear Research, 05-400 Otwock-\'Swierk, Poland \\
$^{b}$ High Energy Physics Division, National Centre for Nuclear Research, 05-400 Otwock-\'Swierk, Poland \\
$^{c}$ Faculty of Physics, Astronomy and Applied Computer Science, Jagiellonian University, 30-348 Cracow, Poland \\
$^{d}$ INFN, Laboratori Nazionali di Frascati, 00044 Frascati, Italy \\ 
$^{e}$ Institute of Physics, Maria Curie-Sk\l{}odowska University, 20-031 Lublin, Poland \\
$^{f}$ Faculty of Physics, University of Vienna, 1090 Vienna, Austria \\
$^{g}$ Department of Physics, College of Education for Pure Sciences, University of Mosul, Mosul, Iraq
}
}

\maketitle
\begin{abstract}
Positron Annihilation Lifetime Spectroscopy (PALS) has shown to be a powerful tool to study the nanostructures of porous materials. Positron Emissions Tomography (PET) are devices allowing imaging of metabolic processes e.g. in human bodies. A newly developed device, the J-PET (Jagiellonian PET), will allow PALS in addition to imaging, thus combining both analyses providing new methods for physics and medicine. In this contribution we present a computer program that is compatible with the J-PET software. We compare its performance with the standard program LT 9.0 by using PALS data from hexane measurements at different temperatures. Our program is based on an iterative procedure, and our fits prove that it performs as good as LT 9.0. 
 
\end{abstract}
  
\section{Introduction}

Nowaday, the decay of the bound state of an electron and a positron, a positronium, could be used in characterizing nanostructures of materials or imaging metabolic processes e.g. in human bodies. The first one is achieved by a technique know as PALS (Positron Annihilation Lifetime Spectroscopy) \cite{Schrader}, \cite{Coleman}, the second one by a Positron Emission Tomograph (PET)  \cite{Van}, \cite{Slomka}, \cite{Karp}, which is used in many hospitals. Joining these two techniques can extend diagnosis possibility of standard PET scan, and possibly improve the sensitivity to diagnose different stages of cancers \cite{Patent0}. \\
\\
J-PET is a newly designed \cite{Prof0} and constructed PET device in Jagiellonian University \cite{Prof01,Raczynski,Smyrski,Alek,Daria,Prof2,Prof3, Szymon}, which due to its original idea of using cheap and fast plastic scintillators and dedicated fast electronics \cite{Palka}, \cite{Korcyl} can in future compete with commercial PET scanners. Furthermore, superior time resolution in J-PET can be used in implementing positronium lifetime measurements, where time measurements are crucial in recovery of lifetime distribution. \\
\\
The idea how to obtain the PALS spectra using PET scanners were introduced in \cite{Patent0}, \cite{Patent}, \cite{Prof1}, \cite{Jasin}. In this paper we present a new procedure to do a PALS analysis by a software compatible with the software used in J-PET \cite{Krzem}. A compatible software would potentially accelerate the analysis process fo data gathered by J-PET detector. The precision and reliability of our new developed software is compared with the broadly used program LT 9.0 \cite{Kansy}.  

\section{Analysis in PALS}

Positronium lifetime in material, in ideal situation follows an exponential decay law $f(t, \tau) = \frac{1}{\tau} \exp \left( - \frac{t}{\tau } \right) $, where $\tau$ is the mean lifetime. However in the measurement it is distorted by the resolution function of the apparatus that is used to gather the PAL spectra. Assuming that resolution impact can be described by the sum of a finite number of Gauss distributions $g(t, t_0, \sigma) = \frac{1}{\sqrt{2 \pi}\sigma}\exp \left( - \frac{\left( t-t_0 \right) ^2}{2 \sigma ^2} \right)$, lifetime component measured in PALS will then follow distribution given by convolution of exponential and Gauss distributions. Because exponential distribution function is only differentiable from zero to infinity one can use one-sided convolution to obtain simplified final form shown in following equation \eqref{Conv} :
\[
\left( f \ast g \right) (t) \overset{df}= \int _0 ^t f(s) g(t-s) ds =
\]
\begin{equation}\label{Conv}
=
\frac{1}{2 \tau} \exp \left( \frac{\sigma ^2}{2 \tau ^2} - \frac{t - t_0}{\tau} \right) \left( \text{erf} \left( \frac{t - t_0 - \frac{\sigma ^2}{\tau}}{\sqrt{2} \sigma} \right) - \text{erf} \left( \frac{- t_0 - \frac{\sigma ^2}{\tau}}{\sqrt{2} \sigma} \right) \right).
\end{equation}
In the material with multiple components with different mean lifetimes, PAL spectra can be fitted by a function $F(t)$ composed of a linear combination of \eqref{Conv} and distributions assuming \textit{n} lifetimes and \textit{m} Gauss resolution components \eqref{final}: 
\begin{equation}\label{final}
F(t) = \sum _{i=1}^{n}\sum _{j=1}^{m} I_{i,j} \left( f( t, \tau _i) \ast g( t, t_0^{(j)}, \sigma ^{(j)}) \right) (t) + y_0 ,
\end{equation}
where $ \text{erf()} $ is the Gauss error function and $y_0$ is the background.\\
$I_{i,j}$ denotes the intensity, so the fraction of a given component of the whole spectrum is always greater than zero.
In that context iterative procedure was developed, programmed in C++ based on ROOT libraries, where each iteration releases gradually more parameters initially fixed. The program handels data in the form of histogram and some initial conditions may be defined by the user. Firstly the background $y_0$ is estimated as the mean of the 50 last points of the histogram, then the initial conditions are taken under consideration. These condition are as follows: (i) number of lifetime components, (ii) their initial mean lifetime and intensity, (iii) degree of freedom (\quotedblbase \textit{f}\textquotedblright $ $, \quotedblbase \textit{p}\textquotedblright $ $ or \quotedblbase \textit{nf}\textquotedblright $ $ explained hereafter in detail), (iv) number of resolution components, (v) their initial width $\sigma$ and shift $t_0$. The last initial parameter given by the user is the number of iteration that should be performed, by default it is set to 3 iterations.\\
\\
The degree of freedom determines how program will be treating given variable in the process of fitting. It can accept \quotedblbase \textit{f}\textquotedblright $ $ as the rigorously fixed, so the lifetime and intensity of given components can not change at all. It is a~proposed option for annihilation component in source, which in principal is known from previous PALS studies. The second option is \quotedblbase \textit{p}\textquotedblright $ $, which fixes component parameters in the first iteration, but in next steps of fitting it is allowed to change the lifetime value up to 5 percent value taken from the previous step. In that option there is no boundary on the intensity value. The last option is \quotedblbase \textit{nf}\textquotedblright $ $, where no boundaries for the next steps are defined. It is also possible to fix a ratio between intensities for some lifetime components, what in general is used to fix the \textit{o}-Ps to \textit{p}-Ps intensity ratio as 3:1.\\
\begin{center}
\includegraphics[scale=0.25]{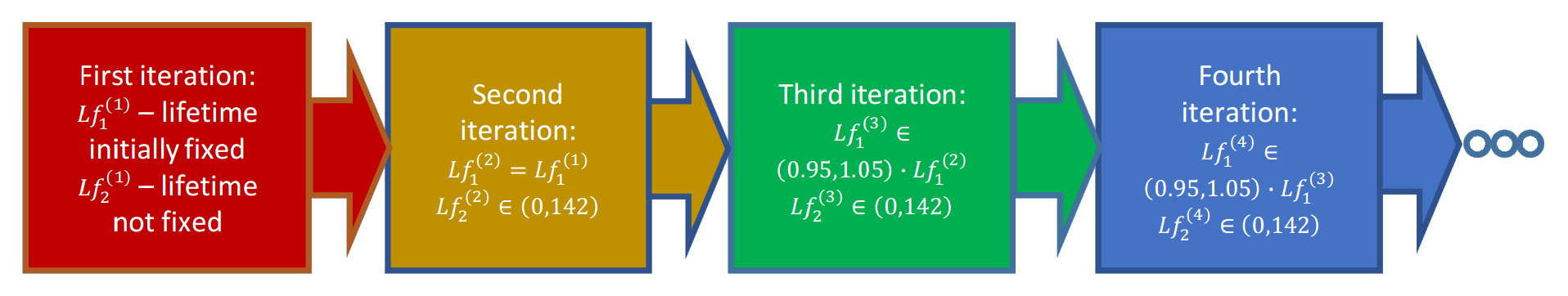}
\\
\begin{small}
\textit{Fig. 2.1. Scheme of the iteration procedure.}
\end{small}
\end{center}
In the first step of the run, the program processes the initial parameters defined by the user and fits a spectrum with fixed lifetime and intensity for components with \quotedblbase \textit{f}\textquotedblright $ $, and \quotedblbase \textit{p}\textquotedblright $ $ options. Then in the second iteration, new initial parameters are set based on the results from the first iteration and fit the spectrum with fixed lifetime and intensity for components with \quotedblbase \textit{f}\textquotedblright $ $ option, but with changed variability of lifetime for \quotedblbase \textit{p}\textquotedblright $ $, so the lifetime can change up to 5\% from initial value. Then the further iterations are performed similarly, until the desirable number of iteration is reached. The whole idea is shown in Fig. 2.1.

\section{Comparison with LT of hexane sample}

To check the reliability of our developed program, it is important to compare the results of the fit with the known and already approved programs like for example LT developed by J. Kansy \cite{Kansy}, PALSfit \cite{PALSfit} and PAScual \cite{PAScual}. Data that were analyzed came from a hexane PALS temperature scan, measured in the PALS laboratory at UMCS, Lublin. Data were collected using source in Kapton foil. Iteration method to fit the data is specified in previous section. Four lifetimes components ($n = 4$) and one resolution component ($m=1$) were applied in the fitting procedure. The fraction of \textit{ortho}-Positronium (\textit{o}-Ps) to \textit{para}-Positronium (\textit{p}-Ps) probability of annihilation was fixed to 3:1. The annihilation in source was fixed by \quotedblbase \textit{f}\textquotedblright $ $ at a mean lifetime of 382 ps with the intensity of 10 \%. The other components were treated as free \quotedblbase \textit{nf}\textquotedblright $ $.
\begin{center}
\includegraphics[scale=0.5]{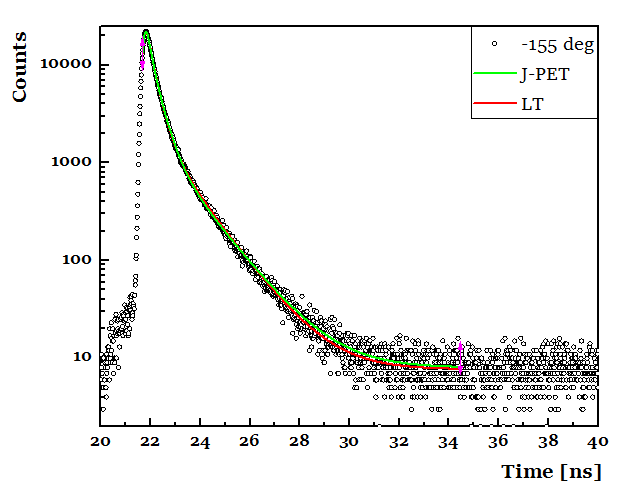}
\\
\begin{small}
\textit{Fig. 3.1. PALS spectrum of hexane measurement in -155 $^o$}C \textit{with fits from LT software and software developed for J-PET. Pink vertical lines indicate the range of fit. Results from fit are shown in Tab. \ref{tab1}}
\end{small}
\begin{table}[h]
\caption{Results from fit of measurement in -155$^o$C}\label{tab1}

\begin{tabular}{|c|c|c|c|c|}
\hline
 & \multicolumn{2}{|c|}{ \multirow{2}{*}{ \textbf{LT} } } & \multicolumn{2}{|c|}{\textbf{Software developed}}\\
  & \multicolumn{2}{|c|}{} & \multicolumn{2}{|c|}{\textbf{by J-PET}}\\
\hline
\hline
Background & \multicolumn{2}{|c|}{7.67 (22)} & \multicolumn{2}{|c|}{8.06 (31)}\\
\hline
FWHM of  & \multicolumn{2}{|c|}{ \multirow{2}{*}{ 0.225 (16) } } & \multicolumn{2}{|c|}{ \multirow{2}{*}{ 0.227 (13) } }\\
Gauss [ns] &  \multicolumn{2}{|c|}{} &  \multicolumn{2}{|c|}{} \\
\hline
Offset of & \multicolumn{2}{|c|}{ \multirow{2}{*}{ 21.52 } } & \multicolumn{2}{|c|}{ \multirow{2}{*}{ 21.68 } } \\
Gauss [ns] &  \multicolumn{2}{|c|}{} &  \multicolumn{2}{|c|}{}  \\
\hline
\hline
 & Lifetime [ns] & Intensity [\%] & Lifetime [ns] & Intensity [\%]\\
\hline
Component 1 & \multirow{2}{*}{0.382} & \multirow{2}{*}{10} &  \multirow{2}{*}{0.382}  & \multirow{2}{*}{10}  \\ 
(source) & & & & \\
\hline
Component 2 & \multirow{2}{*}{1.253 (55)} & \multirow{2}{*}{25.17 (19)} &  \multirow{2}{*}{1.193 (06)}  & \multirow{2}{*}{26.34 (20)}  \\ 
(\textit{o}-Ps) & & & & \\
\hline
Component 3 & \multirow{2}{*}{0.130 (45)} & \multirow{2}{*}{8.39 (06)} &  \multirow{2}{*}{0.137 (08)}  & \multirow{2}{*}{8.78 (07)}  \\ 
(\textit{p}-Ps) & & & & \\
\hline
Component 4 & 0.311 (27) & 56.44 (25) & 0.306 (01) & 54.78 (03)  \\
\hline
\end{tabular}
\end{table}
\end{center}
\begin{center}
\includegraphics[scale=0.5]{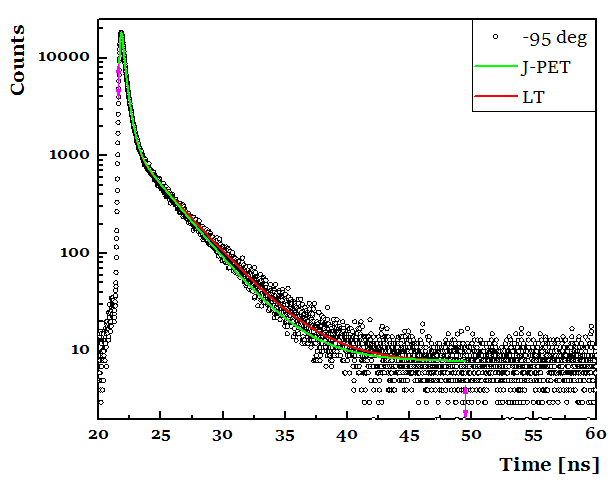}
\\
\begin{small}
\textit{Fig. 3.2. PALS spectrum of hexane measurement in -95 $^o$}C \textit{with fits from LT software and software developed for J-PET. Pink vertical lines indicate the range of fit. Results from fit are shown in Tab. \ref{tab2} }
\end{small}
\begin{table}[h]
\centering
\caption{Results from fit of measurement in -95$^o$C}\label{tab2}
\begin{tabular}{|c|c|c|c|c|}
\hline
 & \multicolumn{2}{|c|}{ \multirow{2}{*}{ \textbf{LT} } } & \multicolumn{2}{|c|}{\textbf{Software developed}}\\
  & \multicolumn{2}{|c|}{} & \multicolumn{2}{|c|}{\textbf{by J-PET}}\\
\hline
\hline
Background & \multicolumn{2}{|c|}{7.75 (23)} & \multicolumn{2}{|c|}{7.95 (27))}\\
\hline
FWHM of  & \multicolumn{2}{|c|}{ \multirow{2}{*}{ 0.226 (32) } } & \multicolumn{2}{|c|}{ \multirow{2}{*}{  0.227 (15) } }\\
Gauss [ns] &  \multicolumn{2}{|c|}{} &  \multicolumn{2}{|c|}{} \\
\hline
Offset of & \multicolumn{2}{|c|}{ \multirow{2}{*}{ 21.52 } } & \multicolumn{2}{|c|}{ \multirow{2}{*}{  21.68 } } \\
Gauss [ns] &  \multicolumn{2}{|c|}{} &  \multicolumn{2}{|c|}{}  \\
\hline
\hline
 & Lifetime [ns] & Intensity [\%] & Lifetime [ns] & Intensity [\%]\\
\hline
Component 1 & \multirow{2}{*}{0.382} & \multirow{2}{*}{10} &  \multirow{2}{*}{0.382}  & \multirow{2}{*}{10}  \\ 
(source) & & & & \\
\hline
Component 2 & \multirow{2}{*}{3.063 (83)} & \multirow{2}{*}{33.09 (19)} &  \multirow{2}{*}{ 2.792 (08)}  & \multirow{2}{*}{ 33.18 (09)}  \\ 
(\textit{o}-Ps) & & & & \\
\hline
Component 3 & \multirow{2}{*}{0.120 (75)} & \multirow{2}{*}{11.03 (07) } &  \multirow{2}{*}{ 0.124 (05)}  & \multirow{2}{*}{ 11.06 (04) }  \\ 
(\textit{p}-Ps) & & & & \\
\hline
Component 4 & 0.362 (32)  & 45.87 (25) & 0.352 (01)  & 45.66 (65)  \\
\hline
\end{tabular}

\end{table}
\end{center}
Exemplary PALS spectra with fits in LT and software developed for J-PET from measurements in -155$^o$C and -95$^o$C of hexane are shown in Fig. 3.1 and 3.2, respectively. Comparison of all measurements for various temperatures are available in Fig. 3.3. One can observe that the results obtained using both programs are very similar, and the pattern in the scan is preserved in both fitting programs showing a phase transition at around -100 $^o$C.

\begin{center}
\includegraphics[scale=0.35]{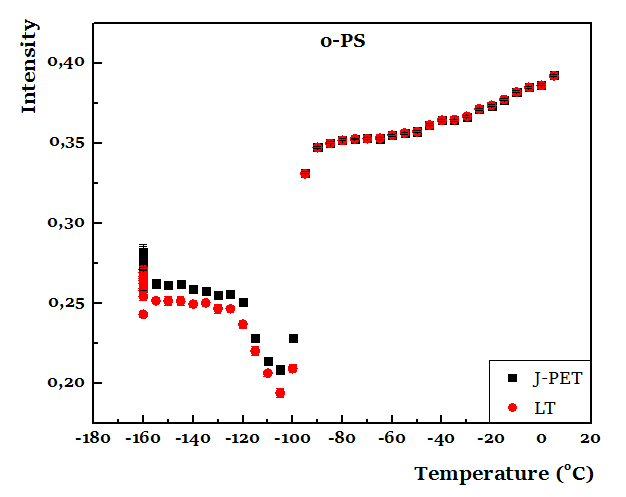}
\includegraphics[scale=0.35]{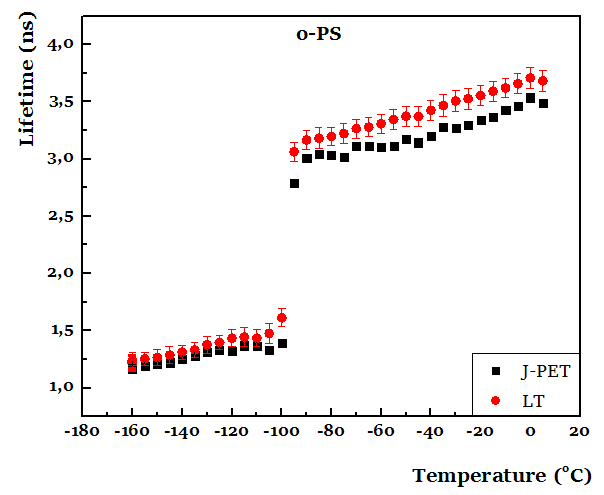}\\
\begin{small}
\textit{Fig. 3.3. Results of the hexane scan shown as dependence of \textit{ortho}-Positroinum lifetime and intensity over temperature.  In the left panel, data from LT and J-PET software above -100 $^o$}C\textit{ are superimposed. }
\end{small}
\end{center}

\section{Conclusions and perspectives}

A new analysis program for the PAL spectra is proposed. The program is based on the iterative procedure compatible with the J-PET tomograph analysis software. The program was validated on the hexane PAL spectra. The obtained results are consistent with results of analysis performed using well established software LT 9.0. \\
\\
It is important to stress that the proposed program is capable in analysing lifetime spectra from conventional and digital spectrometers. It can be also applied for future morphometric images produced by J-PET detector, where the positronium lifetime measured in each voxel would bring additional information in PET scan.

\section*{Acknowledgements}
We acknowledge the support by the National Science Centre through the grant No. 2016/21/B/ST2/01222, and by the Ministry for Science and Higher Education through the grant No. IA/SP/01555/2016. B.C.H. - the Austrian Science Fund (FWF-P26783).



\end{document}